\documentclass[preprint,aps,prd,showpacs,nofootinbib,preprintnumbers,superscriptaddress]{revtex4-1}
\usepackage{hyperref}
\usepackage{mathtools,amsfonts,amssymb}
\usepackage{amsmath,amsthm}
\usepackage{graphicx}
\usepackage{color}

\begin{document}
\title{Inverse-Chirp Imprint of Gravitational Wave Signals in Scalar Tensor Theory}

\author{Chao-Qiang Geng}
\email[Electronic address: ]{geng@phys.nthu.edu.tw}
\affiliation{School of Fundamental Physics and Mathematical Sciences\\Hangzhou Institute for Advanced Study, UCAS, Hangzhou 310024, China}
\affiliation{International Centre for Theoretical Physics Asia-Pacific, Beijing/Hangzhou, China}
\affiliation{Department of Physics,
	National Tsing Hua University, Hsinchu 300, Taiwan}
\author{Hao-Jui Kuan}
\email[Electronic address: ]{guanhauwzray@gmail.com}
\affiliation{Department of Physics,
	National Tsing Hua University, Hsinchu 300, Taiwan}
\affiliation{Theoretical Astrophysics, IAAT, University of Tubingen, Germany}
\author{Ling-Wei Luo}
\email[Electronic address: ]{lwluo@gate.sinica.edu.tw}
\affiliation{Institute of Physics, Academia Sinica, Taipei 11529, Taiwan}

\begin{abstract}
The scalar tensor theory contains a coupling function connecting the quantities in
the Jordan and Einstein frames, which is constrained to guarantee a transformation rule between frames.
We simulate the supernovae core collapse with different choices of coupling functions defined over the viable region of the parameter space and find that a generic inverse-chirp feature of the gravitational waves in the scalar tensor scenario.
\end{abstract}

\maketitle
\begin{section}{Introduction}
	The searches for gravitational waves (GWs) by the kilometric-size laser-interferometer systems, such as 
	the Laser Interferometric Gravitational wave Observatory (LIGO) in  US, 
	Virgo in Italy
	and the KAmioka GRAvitational Wave Detector (KAGRA) in Japan,
	have been initiated to test various gravitational theories.
	Particularly, the gravity effects in the strong-field regime 
	can be verified through the observations, where the underlying gravity theory may
	deviate from General Relativity (GR).
	In practice, several alternative theories of gravity have been proposed. 
	Among them, the scalar tensor (ST) theory is the most natural extension to GR,
	in which gravity can be mediated by a scalar field in addition to the metric one.
    This additional field introduces the spontaneous scalarization phenomenon within the gravitation field of  neutron stars~\cite{Damour:1993hw}, 
	which is a non-perturbative deviation from GR.
	In a recent work~\cite{Sperhake:2017itk}, 
	it has been shown that one can observe the presence of such phenomenon in the supernova
	core collapse scenario as well.
	
	Furthermore, it has been proved by Damour and Esposito-Far\`ese~\cite{Damour:1992we,Damour:1993hw} 
	that under the assumption that there exists a transformation  between Jordan and Einstein frames, 
	a two-parameter family of the ST theory is sufficient to parametrize the most general post-Newtonian
	deviations from GR with the nonperturbative strong-field effects. 
	Many works have also been done along this direction~\cite{Novak:1997hw,Novak:1998rk,Barausse:2012da,Berti:2013gfa,Llinares:2013qbh,Sperhake:2017itk}. However, several problems related to the assumption
	 have been discussed in the literature~\cite{Faraoni:2004is,Flanagan:2004bz,Faraoni:2006fx,Brans:2005ra,Jarv:2015kga,Ishak:2018his,Geng:2020ftu}.
	The transformation between two frames includes a Weyl transformation of metric and
	a redefinition of the scalar field, $d\phi/d\varphi$.
	It has been demonstrated in~\cite{Jarv:2015kga,Geng:2020ftu} that there is a criterion concerning
 the scalar field redefinition,
	which leaves a constraint on the parameter space of the coupling function
	$\alpha(\varphi,\alpha_0,\beta_0)$ defined in the Einstein frame~\cite{Geng:2020ftu}.
	In light of such criterion/constraint, it is prospective to further constrain the ST theory 
	with the signals of GWs.
	To capture the features of those signals, we consider the stellar
	core collapse systems in the massive ST  scenario.
	
	In this work, we first express how the aforementioned criterion manifests itself 
	as a constraint on the parameter space. We then  numerically simulate 
	the supernovae core collapse in the viable region of the parameter space by
	using the code  in \cite{Sperhake:2017itk} to study the profile of the genuine 
	strong-field effects. 
	
	The paper is organized as follows. 
	In Sec.~II, we briefly introduce the theoretical framework of the ST theory by concentrating on
	  the constraint on the scalar field. 
	The numerical  simulations of the  supernovae core collapse are represented in Sec.~III.
	Our conclusions are given in Sec.~VI.
	
\end{section}

\section{Scalar Tensor Theory}
The ST theory can be formulated in both Jordan and Einstein frames, 
which are conformally related. In the Jordan frame, the action takes the form
\begin{align}\label{JD action}
	S = \int d^{4}x\, \frac{\sqrt{-g}}{16\pi G} \bigg( F(\phi)R - \frac{\omega(\phi)}{\phi} 
	g^{\mu \nu} \partial_{\mu}\phi \partial_{\nu}\phi -U(\phi)\bigg)
	+S_{m}[\psi_{m}, g_{\mu \nu}],
\end{align}
where $F(\phi)$ and $\omega(\phi)$ are the regular coupling functions of the scalar field $\phi$, 
and $S_{m}$ corresponds to the action of ordinary matter. 
It has been revealed since the original Brans-Dicke paper appeared~\cite{Brans:1961sx,Dicke:1961gz}
that another formulation of the theory
is possible. Through a Weyl transformation
\begin{align}
	g_{\mu\nu} = A(\phi)^{2}g^{\star}_{\mu\nu},
\end{align}
and a redefinition of the scalar field \cite{Geng:2020ftu}
\begin{align}\label{scalar redefine}
\frac{d\varphi}{d\phi}
\coloneqq \pm \sqrt{\frac{3(F_{,\phi})^{2}}{4F^{2}}
	+\frac{\omega}{2\phi F}},
\end{align}
one can recast the theory into the so-called Einstein frame with the action, given by
\begin{align}\label{Eq:Einstein action}
S = \int d^{4}x\, \frac{\sqrt{-g^{\star}}}{16\pi G} \bigg(R^{\star} 
- 2g^{\star\mu\nu}\partial_{\mu}\varphi \partial_{\nu}\varphi 
- 4V(\varphi) \bigg) +S_{m}[\psi_{m}, A^{2}g^{\star}_{\mu \nu}],
\end{align}
where $g^{\star}_{\mu\nu}$ is the transformed metric, 
$A(\phi)$ is the coupling function defined by $F=A^{-2}$, 
and $V(\varphi) \coloneqq A^{4}U(\phi)/4$. 
As a result, the field equations are the usual Einstein ones
with the scalar field as a source together with an equation of motion of the scalar field, namely
\begin{subequations}\label{Eq:EoM of g and varphi}
	\begin{align}
	&R^{\star}_{\mu \nu} = 8\pi G \bigg( T^{\star}_{\mu \nu} - \frac{1}{2}T^{\star}g^{\star}_{\mu \nu}\bigg)
	+ 2\partial_{\mu}\varphi \partial_{\nu}\varphi 
	+ 2Vg^{\star}_{\mu \nu}
	\label{Eq:EoM of metric},   \\[0.3em]
	&\square^{\star}\varphi 
	= -4\pi G \alpha(\varphi)T^{\star} + \frac{dV}{d\varphi}, 
	\label{Eq:EoM of varphi}
	\end{align}
\end{subequations}
where 
\begin{align}\label{Def alpha}
\alpha(\varphi) \coloneqq \frac{d\ln A}{d\varphi}
= -\frac{1}{2F}\frac{d \phi}{d \varphi}\frac{d F}{d \phi},
\end{align}
and 
$T^{\star} := g^{\star \mu \nu}T^{\star}_{\mu \nu}$ with the stress energy tensor
\begin{align}
T^{\star}_{\mu \nu} 
\coloneqq \frac{-2}{\sqrt{-g^{\star}}} \frac{\delta S_{m}}{\delta g^{\star\mu \nu}}. 
\end{align}

In the mathematical viewpoint,  having the scalar field $\varphi$ in the Einstein frame to be viable
 in the Jordan frame, 
one should be able to  represent $\varphi$ as a function of $\phi$. 
Subsequently, the existence of $\phi(\varphi)$  indicates that \cite{Geng:2020ftu}
\begin{align}\label{const rank thm}
	\frac{d\phi}{d\varphi} \ne 0
	\quad\text{or}\quad
	\frac{d\varphi}{d\phi} \ne 0.
\end{align}
Hence, the solution to the scalar equation in the Einstein frame must satisfy \eqref{const rank thm}.
Otherwise, it is not a solution to the scalar equation in the Jordan frame.

In this work, we adopt the conformal factor $A(\varphi)$ as discussed in~\cite{Damour:1993hw,Damour:1996ke,Novak:1997hw,Novak:1998rk}, given by
\begin{align}\label{coupling func}
	\ln A=\alpha_{0}(\varphi -\varphi_{0})+ \frac{1}{2}\beta_{0}(\varphi-\varphi_{0})^{2},
\end{align}
where $\varphi_{0}$ is the asymptotic value of $\varphi$ at spatial infinity.
The constants $\alpha_{0}$ and $\beta_{0}$ are defined as 
\begin{subequations}
	\begin{align}
	&\alpha_{0} \coloneqq \alpha ( \varphi_{0} ), \\[0.3em]
	&\beta_{0} \coloneqq \frac{d\alpha}{d\varphi}( \varphi_{0} ).
	\end{align}
\end{subequations}
The coupling function  in~\eqref{coupling func} leads to 
\begin{align}\label{Eq:alpha}
	\alpha = \alpha_{0}+\beta_{0}(\varphi-\varphi_{0})
\end{align}
and
\begin{align}
	\ln F=-2 \alpha_{0}(\varphi-\varphi_{0}) - \beta_{0}(\varphi-\varphi_{0})^{2}.
\end{align}
By substituting \eqref{Eq:alpha} into \eqref{Eq:EoM of varphi}, one obtains the equation of motion
\begin{align}\label{wave eq}
\square^{\star}\varphi 
= -4\pi G\alpha_{0}T^{\star} + m_{\text{eff}}^{2}\varphi,
\end{align}
where $m_{\text{eff}}^2$ is the square of the effective mass for $\varphi$, defined by
\begin{align}\label{eff_mass}
	m_{\text{eff}}^2 \coloneqq -4\pi G \beta_0 \varphi T^{\star} + \frac{dV}{d\varphi}.
\end{align}

In~\cite{Damour:1993hw, Damour:1996ke}, Damour and Esposito-Farése
described the dramatic deviation from GR for some specific values of the coupling constants, 
dubbed as ``spontaneous scalarization.''
To trigger this sudden behavior of the scalar field, $\beta_{0}$ should be smaller than a specific value,
which is $-4.35$ for the static neutron stars,
while it would increase a bit but still negative~\cite{Doneva:2013qva}
for the rotating ones.
In this study, we set that
$\alpha_{0} > 0$ and $\beta_{0} < 0$.

The non-vanishing property of \eqref{const rank thm} 
together with the definition \eqref{Def alpha}
implies that the parameter
$\alpha$ can never be zero, resulting in that there is a
critical value for $\varphi$ by \eqref{Eq:alpha}, 
denoted as  $\varphi_{c}$ \cite{Geng:2020ftu},
i.e. 
\begin{align}\label{critical value}
\varphi \ne \varphi_{c} \coloneqq -\frac{\alpha_{0}}{\beta_{0}}.
\end{align}
This shows that the solution space of the scalar field in the Einstein frame is divided
into two disconnected branches by the value of $\varphi_{c}$ with $\varphi=\varphi_{c}$ to be a
non-crossing line for  the scalar field in the Einstein frame.

The metric in the Jordan frame  is partially determined by  $\varphi$, which is a field in the 
Einstein frame. 
Technically speaking, once one ensures that the signals from the simulation are reversible to the 
Jordan frame,
the measurable amplitude of the scalar signal can be expressed as
\begin{align}
	h_s = h_{B} - h_{L},
\end{align}
where 
\begin{align}
	h_{B} = 2 \alpha_{0}\varphi
\end{align}
and 
\begin{align}
	h_{L} = \bigg( \frac{\omega_{comp}}{\omega} \bigg)^{2} h_{B},
\end{align}
which are the breathing and longitudinal modes of GWs, respectively \cite{Will:2014kxa}.
The amplitude of the longitudinal mode $h_L$
is proportional (up to a sign) to $h_{B}$, and the coefficient
depends on the Compton length of the scalar field, $\omega_{comp}\coloneqq m_{\text{eff}}/ \hbar$,
hence to the effective mass term of $\varphi$.
As a result, $h_s$ is clearly proportional to $\varphi$. 
We shall show that this amplitude has an inverse-chirp evolution
along with $\varphi$, which is the key profile of the additional modes of GWs in 
the massive ST theory.
Practically,
the GW signals from the stellar collapse in the theory are monochromatic soon after the emission, 
which are likely detectable with the ground-based GW detectors \cite{Sperhake:2017itk}.

\section{Numerical Simulations}
We follow the method used in \cite{Sperhake:2017itk}.
The supernovae core collapse is considered with the spherically symmetric metric of the form
\begin{align}
	ds^{2} = -F\alpha^{2} dt^{2} + FX^{2} dr^{2} + rd\Omega^2,
\end{align}
where all metric functions depend only on the coordinates $r$ and $t$.
We take  matter as a perfect fluid,
whose energy-momentum tensor in the Jordan frame can be expressed in  the spherical coordinate as
\begin{align}
	T_{\mu\nu} = \rho H u_{\mu}u_{\nu} + pg_{\mu\nu},
\end{align}
where $\rho$, $p$ and $u^{\mu}$ are the energy density, pressure and  4-velocity
of matters, respectively. The enthalpy is defined by 
\begin{align}
H = 1 + \epsilon + \frac{p}{\rho},
\end{align}
where $\epsilon$ is the internal energy.
The quantities are connected to those in the Einstein frame, labeled  with
the asterisk, via the relations
\begin{subequations}\label{17}
	\begin{align}
		&\rho = A^{-4}(\varphi)\rho^{\star}, \\
		&p = A^{-4}(\varphi) p^{\star}, \\
		&u_{\mu} = A(\varphi) u_{\mu}^{\star}.
	\end{align}
\end{subequations}
The field equations are solved numerically by utilizing 
the modification of the code introduced in
\cite{Sperhake:2017itk}, which is developed from GR1D \cite{OConnor:2009iuz}. 
In the simulation, the hybrid equation of state (EOS) is used
to account for the stiffening of the nuclear
and  model the response of the shocked material
by the forms of 
$p=p_{c} + p_{\text{th}}$ and $\epsilon = \epsilon_{c} + \epsilon_{\text{th}}$
with the thermal effects,
where the cold parts of pressure and  internal energy are given as
\begin{subequations}\label{18}
	\begin{align}
		&p_{c} = K_{1} \rho^{\Gamma_{1}}, \quad 
		\epsilon_{c} = \frac{K_{1}}{\Gamma_{1}-1}\rho^{\Gamma_{1}-1}, 
		  \text{\quad as\quad} \rho \le \rho_{\text{nuc}}\\[0.3em]
		&p_{c} = K_{2} \rho^{\Gamma_{2}}, \quad 
		\epsilon_{c} = \frac{K_{1}}{\Gamma_{2}-1}\rho^{\Gamma_{2}-1} + E_{3},
		  \text{\quad as\quad} \rho > \rho_{\text{nuc}},
	\end{align}
\end{subequations}
where $\rho_{\text{nuc}} = 2 \times 10^{14} \text{g/cm}^{3}$ and  $K_{1} = 4.9345\times 10^{14}$ [cgs] with
$K_{2}$ and $E_{3}$ naturally followed by the continuity. An additional relation to close up
Eqs.~\eqref{17} and \eqref{18} is given by
\begin{align}\label{19}
	p_{\text{th}} = (\Gamma_{\text{th}}-1)\rho \epsilon_{\text{th}}. 
\end{align}
Clearly, we  have three parameters $(\Gamma_{1},\Gamma_{2},\Gamma_{\text{th}})$,
 set to be $(1.3,2.5,1.35)$, to specify the EOS.

As a first study of the influence of the constraint in \eqref{const rank thm},
we will present the results for a specific progenitor of the supernova core
collapse, which is coded as WH20 in \cite{Gerosa:2016fri}, 
along with the density of the atmosphere being 2 $\text{g/cm}^{3}$ outside the progenitor.
Note that our methodology 
can be generalized to all systems.

We consider the action
\begin{align}\label{Eq:model}
S = \int d^{4}x\, \frac{\sqrt{-g^{\star}}}{16\pi G} \bigg(R^{\star} 
- 2g^{\star\mu\nu}\partial_{\mu}\varphi \partial_{\nu}\varphi 
- 2m^{2}\varphi^{2} \bigg) +S_{m}[\psi_{m}, A^{2}g^{\star}_{\mu \nu}],
\end{align}
where the coupling function $A(\varphi)$ is given in \eqref{coupling func}.
The effective mass can be obtained from \eqref{eff_mass}
\begin{align}
	m_{\text{eff}}^2 = -4\pi G\beta_{0}T^{\star} + m^{2}.
\end{align}
If GWs are to be detectable inside the LIGO sensitivity 
window, the mass should be bounded above by $10^{-13}$ eV 
since the low-frequency modes of GWs with $\omega<\omega_{comp}$ will damp out instead of
radiating outward to infinity \cite{Jackson:1998nia}.
In addition, the mass less than $10^{-15}$ eV would not be able to generate the strong scalarization 
and satisfy binary pulsar constraints \cite{Gerosa:2016fri, Ramazanoglu:2016kul} at the same time.
Hence, we fix the mass to be $10^{-14}$ eV hereafter.

In principle, one can define the coupling functions 
by fixing $\alpha_{0}$ and $\beta_{0}$
in \eqref{Eq:alpha} with $\varphi_0=0$ \cite{Sperhake:2017itk,Geng:2020ftu} 
to carry out the simulation. 
%
To illustrate  our results,
we use the set containing pairs $(\alpha_0,\beta_0)$ 
with a constant ratio of $-k$, namely
\begin{align}\label{Eq:a set of const k}
	S_{k} = \big\{ (\alpha_{0},\beta_{0}) \big| \alpha_{0}/\beta_{0} =- k \big\}.
\end{align}
It is clear that $S_{k}$ is defined by a certain critical value. 
One can view the solutions of the scalar field for $(\alpha_0,\beta_0)$ within $S_{k}$ as 
an \emph{one-parameter family} curve by choosing $\beta_{0}$ as the parameter for 
the later analysis. 
In Fig.~1, we show that the amplitudes of the GW signals at the distance of $5 \times 10^{9}$ cm
away from the stellar core are too small by comparing with  the critical value $k=0.05$, 
where $\varphi = k = 0.05$ is a horizontal line far above the signals on the plot. 
One can further notice that the signals for the cases with $\beta_{0}=-2$ and $-4$ are obviously different
from the others since they do not  or barely possess the second twist
before reaching the peak  around $0.3s$. 
This will be explained next.

\begin{figure}
	\centering
	\includegraphics[scale=0.35]{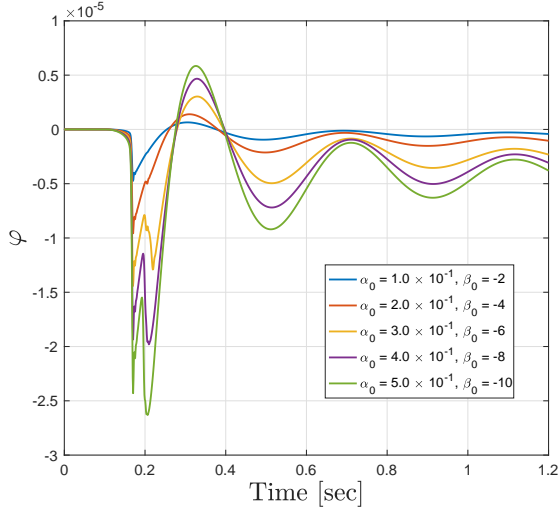}
	\caption{ Waveforms of the scalar field, extracted at r$_{\text{ex}} = 5 \times 10^{9}$ cm
		away from the supernovae core, where the parameters are chosen in the manner that they have
		the same critical value of $\varphi_{c}=k = 0.05$.
	}
\end{figure}
As  introduced in~\cite{roscamead2020core},
the amplitudes of the signals at the wave zone of their propagations with 
different Compton lengths of the scalar field have an approximate universal relation.
If we ignore the source term of $-4\pi G\alpha_{0}T^{\star}$ in \eqref{wave eq}, 
the Compton length would be a function of $\beta_{0}$, so that the signals within the same window
$S_{k}$ have a homologous form. 
This universality is broken due to the presence
of $-4\pi G\alpha_{0}T^{\star}$, which may be measured by the absolute value of 
	the ratio between the coefficients of the zeroth and  first order  terms  in $O(\varphi)$
	on the right hand side of \eqref{wave eq}, given by
\begin{align}
	\delta\coloneqq\frac{-4\pi G\alpha_{0}T^{\star}}{-4\pi G\beta_{0}T^{\star}+m^{2}} 
	=\frac{-k}{1- (4\pi G \beta_{0})^{-1}\gamma}\,,
\end{align}
where $\gamma=m^{2}/T^{\star}$. 
Consequently, 
 we have that  $\delta\approx -k=-0.05$ for $S_{0.05}$ as the linear term dominates,
and $\delta\to 0_{-}$ otherwise.
The behaviors of $\delta$ as a function of $\beta_0$ with several values of $\gamma$  ($\gamma=1$)
are plotted on the left (right) panel of Fig.~2. 
For the signals in Fig.~1,  our simulation on  the right panel of Fig.~2 illustrates that
there are deviations of  $0.79-1.31\%$ from the first order dominating limitation
when $\beta_{0}=-6,-8$ and $-10$, whereas they are about twice even three times as much as 
the  cases with $\beta_{0}=-2$ and $-4$.
From Fig.~2,
one can graphically
see  that the homologous form has been distorted for the later two cases.
\begin{figure}
	\centering
	\includegraphics[scale=0.35]{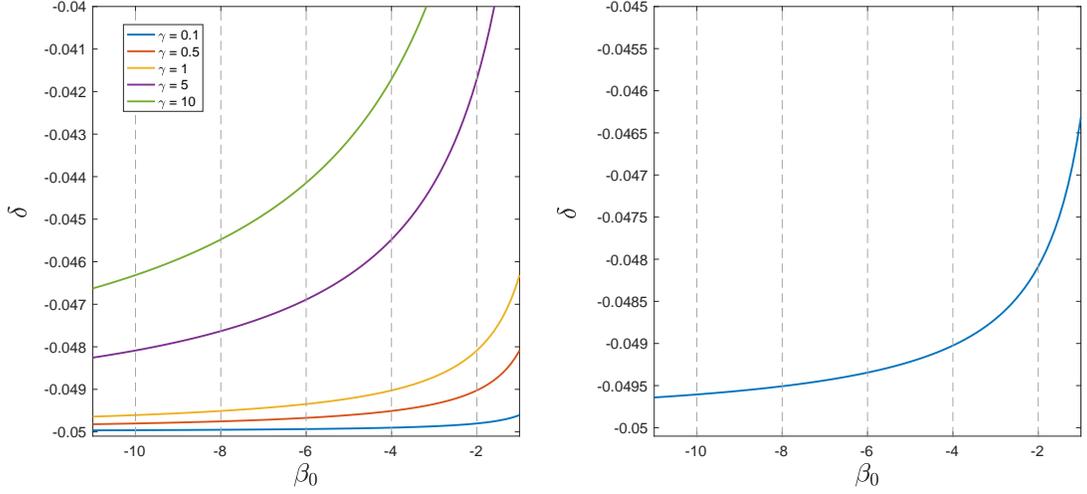}
	\caption{ 
		Behaviors of $\delta$ as a function of $\beta_0$,
		where the left panel represents the cases  in Fig.~1 with $S_{k=0.05}$ 
		by fixing $\gamma$ to be
		$0.1,0.5,1,5$ and $10$, respectively, while the right panel is our simulation with 
		 $\gamma=1$.		 
		For both panels, the vertical dashed lines for
		$\beta_{0}=-2,-4,-6,-8$, and $-10$ are the values for the signals in Fig.~1.}
\end{figure}

For  $S_{1\times 10^{-5}}$  in Fig.~3,
the amplitudes of the GW signals at the same extraction distance are comparable to those at the critical value
$\varphi = k = 1\times 10^{-5}$, 
and hence the constraint is more stringent in this case. 
Any solution
crossing the dashed line $\varphi=k$ will be ruled out.
In the cases shown in Fig.~3, $\delta$ is confined in $0.20-0.26\%$ with respective to 
	$-k=-1\times 10^{-5}$, which is small enough so
that the shapes of the signals do not deviate much.
\begin{figure}
	\centering
	\includegraphics[scale=0.35]{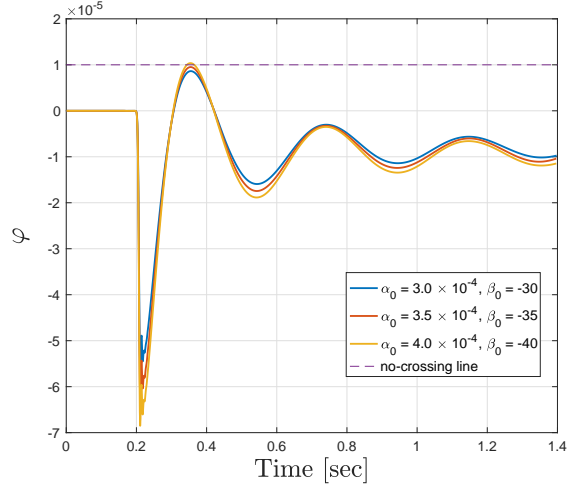}
	\caption{Legend is the same as Fig.~1, but with
		$\varphi_{c}=k = 1\times 10^{-5}$, where the scalar field labeled by $\beta_{0}=-40$ is ruled out
		by the argument in the context.
	}
\end{figure}
%
From Figs.~1 and 3, one can observe that
as $\beta_{0}$ decreases, the peaks of the curves  increase accordingly until 
they touch the non-crossing line. We can define the corresponding parameter as
the critical value of $\beta_{0}$, denoted by $\beta_{c}$.
As a result, there exists a value of $\beta_{c}$ 
such that the peak of $\varphi$ reaches
the value of $\varphi_{c}$.
Consequently, any case with $|\beta_{0}|>|\beta_{c}|$ 
will be forbidden due to its crossing with the line of $\varphi=\varphi_{c}.$

In Fig.~4, we select seven values of $1\times10^{-5}$, $8 \times 10^{-6}$, $7 \times 10^{-6}$, 
$5\times10^{-6}$, $2.5 \times 10^{-6}$, $1\times10^{-6}$, and $5\times10^{-7}$ for $k$ as the
 constraints on $\beta_{0}$.
The parameter space is split into two pieces bounded by the solid curve, in which  the parameters in the left (yellow) region 
are ruled out.
This is to say, the Jordan and Einstein frames do not 
correspond to each other in the shade area in Fig.~4.
We note that the shade area will change for the different initial data/progenitors
as well as EOS.

We now discuss  the  dependence of our results on the hybrid EOS.
The scalar field signal grows in the amplitude during the collapsing phase, while the 
central mass density increases. As the density gets beyond the nuclear density 
$\rho_{\text{nuc}}$, the stellar core undergoes a short period of bounces.
After this phase, the scalar field in the inner core tends to be a static profile, while the scalar radiation 
	is generated outwards. 
	During the process, $\Gamma_{2}$ and $\Gamma_{\text{th}}$ in Eqs.~(\ref{18}) and (\ref{19})
	related to the EOS only affect the wave signal 
	during and after the bounces, so that the influence of the  EOS on the scalar waves is 
	mainly from $\Gamma_{1}$. Explicitly, a larger  $\Gamma_{1}$ would
	result in a more compact core at the moment of the bounce, 
	while
	the stronger
	scalar waves would be released accordingly \cite{Gerosa:2016fri}.  
	This leads to a larger shaded area because the peak of the scalar field reaches the 
	non-crossing line with a smaller $|\beta_{0}|$.
	However, the inverse-chirp feature of
	the scalar radiations remains unchanged as long as one considers the viable parameter space.

\begin{figure}
	\centering
	\includegraphics[scale=0.35]{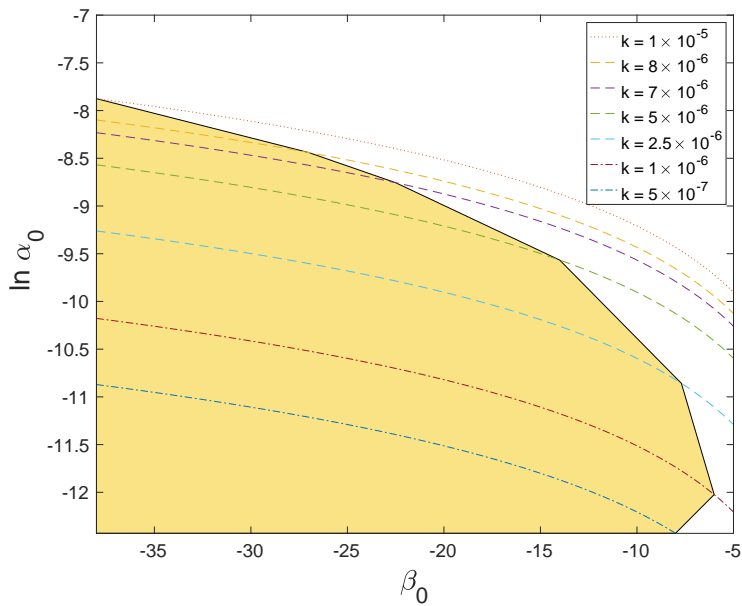}
	\caption{
	Constraints on the coupling parameters $\alpha_{0}$ and $\beta_{0}$ in~\eqref{Eq:alpha} with $\varphi_0=0$ and
	$k=1\times10^{-5}$, $8 \times 10^{-6}$, $7 \times 10^{-6}$, 
	$5\times10^{-6}$, $2.5 \times 10^{-6}$, $1\times10^{-6}$ and $5\times10^{-7}$, respectively,
	where the  shaded region of $(\beta_{0},\ln\alpha_{0})$, given by the critical values $\beta_c$, marks the inviable parameters.
	}
\end{figure}

Moreover, within the same parameter set of $S_{k}$,
the amplitude of the scalar field with a larger $|\beta_{0}|$ or $\alpha_{0}$ is bigger.
The contribution of the scalar field in the Einstein frame to the 
gravitational waveform in the Jordan frame comes through Eq. (5.6), given by~\cite{Damour:1992we}
\begin{align}\label{Eq:contribution in tensor mode}
	2\alpha_{0}\varphi \eta_{\mu\nu},
\end{align}
which implies that the scalar field $\varphi$ affects more as $\alpha_{0}$ increases.
However, to some extend, the solution will touch the non-crossing line and
this  $\alpha_{0}$ puts the upper limit for the detectability of the contribution
of the scalar field in GWs.

\section{Conclusions}
We have simulated the supernovae core collapse and 
found the generic feature of scalar GWs in the ST scenario 
with $V=(1/2)m\varphi^2$ in the Einstein frame, 
which has an inverse-chirp behavior.
We have shown that the ST theory should 
be defined in the viable region for the parameter space 
in order to have the signal to be recognized as GWs in the Jordan frame.

In particular, 
to ensure that one can transform the fields from the Jordan frame to Einstein one, and vice versa,
there is a constraint on the parameter space in the ST theory. 
For the supernovae core collapse, we have illustrated  the 
upper bound on $-\beta_{0}$ for each $S_k$ in \eqref{Eq:a set of const k}.
Using the  bounds  for the different sets of $S_k$,
we have obtained the viable 
region for the parameters in the particular ST theory.
In this area of the parameter space,
we have carried out the numerical simulations with several pairs of $(\alpha_{0}, \beta_{0})$.

Even though the intrinsic amplitude of the scalar field is insensitive to 
$(\alpha_0,\beta_0)$ \cite{Sperhake:2017itk}, 
the measurable scalar signal with the amplitude $h_s$ in terms of $\varphi$
is closely related to the Weyl transformation
associated with the viable region of these parameters.
Furthermore, the constraint of the parameter would affect the contribution of the scalar mode
in the gravitational waveform of the tensor mode, which has been shown in 
\eqref{Eq:contribution in tensor mode}. Therefore, we have demonstrated that the inverse-chirp 
profile of the GW signals is generic so that it can be used as a probe to test the ST theory. 
%

\begin{acknowledgments}
We would like to thank Patrick Chi-Kit Cheong for useful discussions.
The work was partially supported by National Center for Theoretical Sciences
and
Ministry of Science and Technology (MoST-107-2119-M-007-013-MY3 and MoST-108-2811-M-001-598).
\end{acknowledgments}


\end{document}